# Antimony Chalcogenide-based Solid State Sensitizers for Solar Cells: A Forgotten Hero or Low Potential Candidate

Sumanshu Agarwal[1], Harekrishna Yadav[2], and Kundan Kumar[1]

[1]Department of Electronics and Communication Engineering, Institute of Technical Education and Research, Siksha 'O' Anusandhan (Deemed to be University), Bhubaneswar, Odisha—751030, India
[2]Discipline of Mechanical Engineering, Indian Institute of Technology Indore, Indore, Madhya Pradesh— 452020 , India
sumanshuagarwal@soa.ac.in

*Abstract* — The use of stibnite ($Sb_2S_3$) as sensitizers in the solid-state sensitized solar cells received considerable research interest during the transition of the millennium. However, the use of perovskite diminished the research in the field and the potential of antimony chalcogenide ($Sb_2(S,Se)_3$) was not explored thoroughly. Although these materials also provide bandgap tuning like perovskite by varying the composition of S and Se, it is not as popular as perovskite mainly because of the low efficiency of the solar cells based on it. In this paper, we present a landscape of the functional role of various device parameters on the performance of $Sb_2(S,Se)_3$ based solar cells. For the purpose, we first calibrate the optoelectronic model used for the simulation with the experimental results from the literature. The model is then subjected to parametric variations to explore the performance metrics for this class of solar cells. Our results show that despite the belief that open circuit voltage is independent of contact layers doping in proper band aligned sensitized solar cells, here we observe otherwise and the open circuit voltage is indeed dependent on the doping density of the contact layers. Using the detailed numerical simulation and analytical model we further identify the performance optimization map of $Sb_2(S,Se)_3$ based sensitized solar cells.

## I. Introduction

The late 20th century has witnessed rapid research in various solar cell technologies[1,2], including thin-film[3–5] and multijunction[6] solar cells, and also the dye-sensitized solar cell (DSSC) was invented[7] during that time period. Although initially, DSSCs were supposed to have a promising future[8], very soon, it was beyond the horizon because of the lack of reproducibility[9], scalability[10], and stability[11,12] along with the degrading performance[13] and electrode corrosion[14]. In that context, the use of $Sb_2S_3$ in solar cells[15] opened a gateway to solid-state sensitized solar cells (SSSCs, see fig 1 for the schematic) and hence, mitigating the problems of stability, corrosion, and reproducibility[16]. However, the technology did not draw much attention because of substandard performance characteristics[17], and research in the field had been shallow[18]. Contrastingly, exponential growth in the research on perovskite solar cells in the last decade[19,20], owing to their high performance and reproducibility, has hardly been overlooked by anyone[21,22]. On that account, the research on antimony chalcogenide-based SSSCs was not endorsed much, and the technology did not see expected growth.

Nonetheless, a few research groups have reported exciting results in the area[23,24] ranging from bandgap tuning[25] (in the range of 1.1 eV to 1.71 eV) and compositional grading of the active layer[26] (by using $Sb_2(S,Se)_3$) to fabrication of high-efficiency devices[25,27,28] using relatively simple techniques[29]. Despite these encouraging achievements, it is worth noting that the performance loss is very high in these devices (state of the art efficiency ($\eta$) is ~10%[25] as compared to the corresponding SQ limit[30] of ~32%). At this juncture, it is apparent that if the technology is to shine under the sun in the future, then it is important to elucidate the physical mechanisms governing the losses in the device, and strategy for the performance improvement need to be canvased.

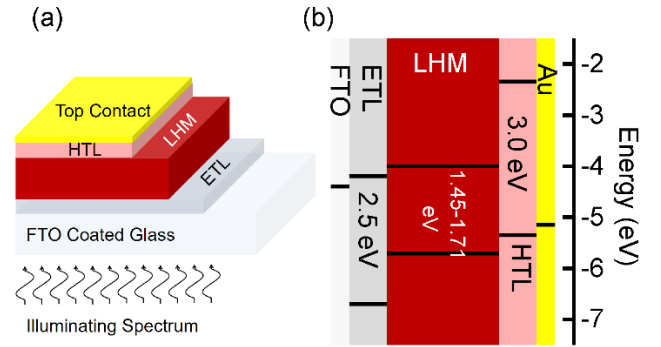

Fig. 1. Schematic and band level alignment of antimony chalcogenide-based sensitized solar cell. (a) Schematic of a typical antimony chalcogenice-based sensitized solar cell. Here, the light harvesting material (LHM) is $Sb_2(S_{1-x}Se_x)_3$. (b) Band level alignment of different materials. Depending on the value of "x" in the range 0-48% in LHM its band gap could be in the range from 1.45 eV to 1.71 eV.

Given that, here, we present a comprehensive study of a) working principle, b) loss mechanisms, and c) performance optimization route for $Sb_2(S_{1-x}Se_x)_3$ based solar cells. Indeed, we provide a modeling framework and detailed numerical simulations calibrated with experimental results from the literature to show that (a) poor lifetime of the charge carriers in the active layers result in a significant loss in short circuit current and open circuit voltage ($V_{OC}$), (b) $V_{OC}$ is further limited by the high interfacial recombination of the charge carriers, and (c) doping density in the contact layers affect $V_{OC}$ by recucing the recombination of the photogenerated charge carriers. Finally, using our simulation results we prescribe engineering solutions for ~11% efficient $Sb_2(S_{1-x}Se_x)_3$-based solar cells (basically the practical efficiency limit for the cases under investigation), which could further be pushed to a higher value by fabricating high-quality $Sb_2(S_{1-x}Se_x)_3$.

Schematic of a typical $Sb_2(S_{1-x}Se_x)_3$ solar cell is shown in Fig. 1 (a). Like any other sensitized solar cells, $Sb_2(S_{1-x}Se_x)_3$ based sensitized solar cells are consist of an active layer or light-harvesting material (LHM, which is $Sb_2(S_{1-x}Se_x)_3$ itself) sandwiched between the electron transport layer (ETL) and hole transport layer (HTL)[18]. While LHM harvests the solar radiation and generates free electron-hole pairs, ETL and HTL ensure the selective collection of the photogenerated charge carriers. Usually, FTO coated glass is used as the substrate, which also acts as n-type contact in the device and Au as back contact is deposited over HTL. We consider the same NIP design (light is incident from the ETL side) in our study. However, the results can readily be extended to PIN configuration by using a proper optical generation profile.

## II MODEL SYSTEM

To elucidate the functional role of various parameters on the performance of SSSCs under investigation, we simulate current-voltage characteristics of the devices under dark and illuminated conditions through self-consistent solutions of drift-diffusion and Poisson's equations[31]. Band level alignments of the different layers used in this study are shown in Fig. 1 (b). Here, we adapt CdS properties for ETL (100 nm thick) and spiro-MeOTAD properties for HTL (100 nm thick). An exponential photo-generation profile of charge carriers[32] along the thickness of the LHM with an effective decay constant ($\alpha_{eff}$) being $10^5$ cm$^{-1}$ (a value close to $\alpha$ in the visible range of the spectrum, where most of the generation takes place[25]) has been assumed. Accordingly, the position-dependent photogeneration rate of the charge carriers in LHM is given by[32]

$$G(x) = G_0 e^{-\alpha_{eff} x}, \qquad (1)$$

where $G_0$ is the photogeneration rate of the charge carriers inside LHM at the ETL/LHM interface. Moreover, we consider all three types of recombination mechanisms viz. radiative recombination, SRH recombination, and Auger recombination along with the interfacial recombination at ETL/LHM and LHM/HTL interfaces. To account for the interfacial recombination, we consider the 4 nm thick interface regions[33] at ETL/LHM and LHM/HTL interfaces (see fig 2a) having very poor SRH lifetime (see table S1 in the supplementary material for the details). Although we adapt most of the material parameters for the different layers from the literature[17,18,25,34,35], transport properties (SRH trapping time and mobility of the charge carriers) for the active layer are estimated using extensive simulations (details are discussed later). It is worth to mention here that instead of forcing the contacts to be ohmic, we use real work functions values for the same, *i.e.* 4.4 eV for FTO[36] and 5.15 eV for Au[25] to take care the Schottky barrier effects at the electrodes/semiconductor interfaces.

## III RESULTS AND DISCUSSIONS

*a) Calibration, working principle, and loss mechanisms:*

Fig 2a shows the equilibrium energy band diagram. We observe a constant electric field of the order of 6200 V/cm in the active layer, which is responsible for sweeping the photogenerated charge carriers and hence, generating current under illumination. Despite the high electric field in the active layer at short circuit, the photogenerated charge carrier collection efficiency is significantly less than unity in the case of $Sb_2(S,Se)_3$ based solar cells[25]. This is because the drift length of the charge carriers ($L_n$ for electrons and $L_p$ for holes) in the active layer, as given by

$$L_\xi = \mu_\xi \tau_\xi E, \qquad (2)$$

governs the collection efficiency under the drift limit[37]. Here, $\mu_\xi$ is the charge carrier mobility, $\tau_\xi$ is the effective lifetime of the charge carrier ($\xi$ is replaced by $n$ for electrons and $p$ for holes), and $E$ is the electric field in LHM. Accordingly, reported poor lifetime of the charge carriers[25] could be one of the reasons for the atrocious collection efficiency in these devices. However, to explore the design schemes for the device performance improvement, it is important to have an accurate estimate for the effective transport parameters (SRH trapping time ($\tau_{SRH}$) and mobility) for the active layer in the device.

To identify the transport parameters, we perform extensive drift-diffusion simulations and compare the current-voltage characteristics with the experimental results. While equation 2 indicates that the lousy collection efficiency of the devices at short circuit point could be due to inept values of lifetime ($\tau$) or mobility ($\mu$) of the charge carriers in the active layer, approximately 50% loss in $V_{OC}$ (0.6–0.7 V) as compared to corresponding limiting values (1.2–1.4 V) confirms that LHM has a very low lifetime of the charge carriers[38]. Indeed we find that $\tau_{SRH}$ for the bulk of the LHM in all the cases under investigation are in the range of tens of picoseconds (comparable to the reported values[25]). Moreover, 4 nm thick interfacial regions at the interfaces of ETL/LHM and LHM/HTL have the value of SRH lifetime less than an attosecond (see table S1 for the details).

Current-voltage characteristics under dark obtained from the simulation for the devices with different LHM using the parameter space discussed above are shown in figure 2b. For all the cases, ideality factor ($n$) was found to be ~1.5 for a broad range of applied bias (except $Sb_2S_3$-based device, where a transition in $n$ from ~1.5 to ~0.8 is observed in the range of $V = 0.3 - 0.7$ V), which indicate SRH dominant recombination in the devices (see fig S1 in the supplementary material for the details). Further, as expected, reverse saturation current density ($J_0$) increases with the decrease in the bandgap of the active layer because of an increase in the generation current ($\sim n_i/\tau$)[39]. The implications of the same are observed in current-voltage characteristics under illumination, where we observe a reduction in $V_{OC}$ with a reduction in the bandgap (fig 2c). Figure 2c compares the simulated and experimental[25] current-voltage characteristics under illumination. The experimental $J - V$ ($J$ is the current normalized w.r.t. active area of the device at the hole collecting electrode) characteristics are reproduced by the numerical simulations. Hence, the results validate the parameter space used for the simulations. Interestingly, we find that short circuit current density ($J_{SC}$) values observed in all the cases are much less than the integrated value of the photogeneration rate of the charge carriers in the active layer (e.g., in the case of $Sb_2S_3$-based solar cell, the collection efficiency at short circuit point is ~87%), which in turn justifies the slope at $J_{SC}$ point.

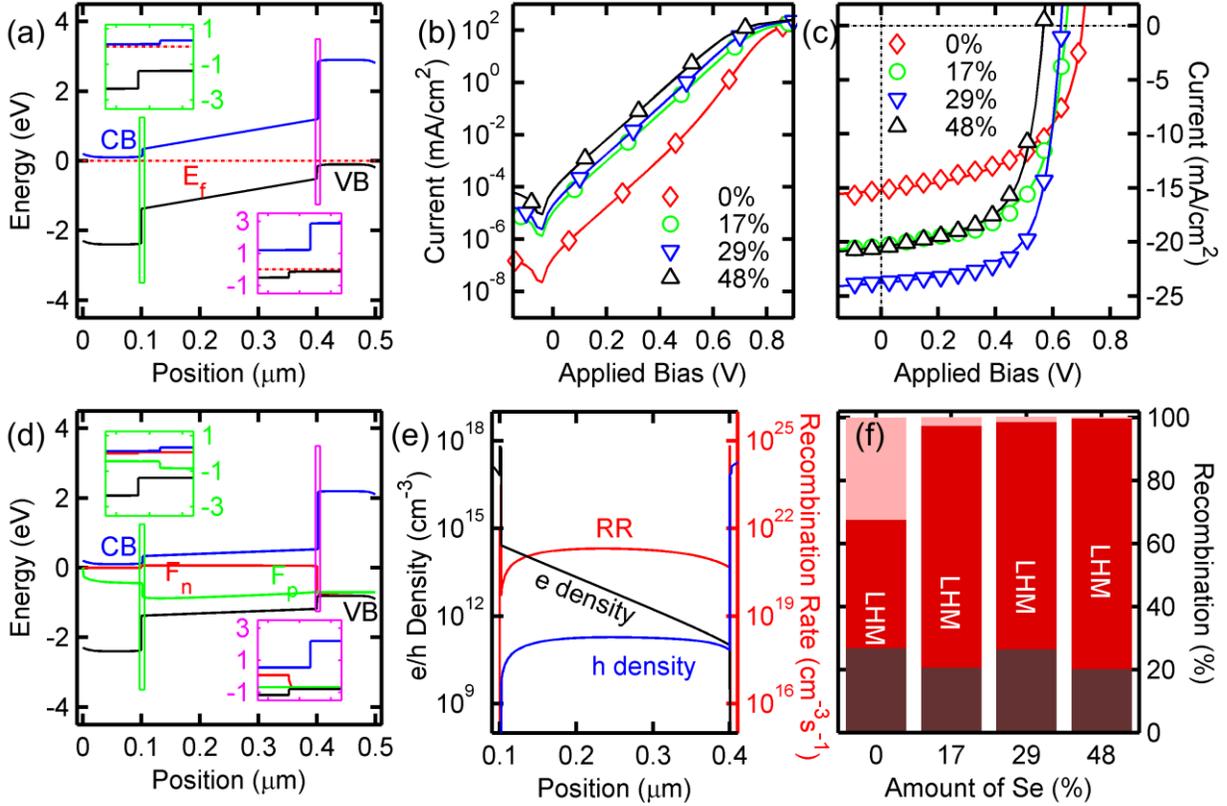

Fig. 2. Analysis of working principle of $Sb_2(S_{1-x}Se_x)_3$ based solar cells. (a) shows the equilibrium band diagram of for $Sb_2S_3$-based solar cell. (b) shows the simulated current voltage characteristics for $Sb_2(S_{1-x}Se_x)_3$ solar cells under dark. The percentage values in the legend are corresponding to "x" in $Sb_2(S_{1-x}Se_x)_3$. (c) shows the simulated current voltage characteristics for $Sb_2(S_{1-x}Se_x)_3$ solar cells under illumination. Here, solid lines indicate the simulation results while experimental results[25] are shown by symbols. (d) and (e) show the band diagram and carrier density and recombination rate at $V_{OC}$. Percentage of recombination in different regions in the device at $V_{OC}$ is shown in (f). The top regions in the bars indicate recombination at LHM/HTL interface, middle regions indicate recombination in the bulk of the active layer, and bottom region indicate recombination at LHM/ETL interface.

To further explore the charge carrier dynamics in the devices under illumination, we plot the band diagram in figure 2d and free charge carrier density and recombination rate in figure 2e, all at an applied bias equal to $V_{OC}$ under illumination condition. In fig 2d, we observe that the slope of the bands is favorable for the collection of photogenerated charge carriers at the desired electrode, despite that $V_{OC}$ is attained. This is mainly due to the recombination of all the photogenerated charge carriers in the active layer. We confirm the same by analyzing the total recombination in the device at $V_{OC}$. Here, we would like to stress the fact that the total recombination of the charge carriers at $V_{OC}$ is not equivalent to $J_{SC}$; in fact, it is equivalent to the integrated photogeneration rate. Furthermore, because of the carrier selective contact layers (see fig S2 in the supplementary material for the details) the current-voltage characteristics under illumination indeed follow the following relation

$$J_l = -q \int_t (G(x) - R(x)) \, dx \qquad (3)$$

where $J_l$ is the normalized current at the hole collecting electrode, $q$ is the elementary charge, $t$ is the thickness of the LHM, $G(x)$ is the photogeneration rate and $R(x)$ is the recombination rate of the charge carriers in the LHM. The analysis of the recombination profile shown in figure 2e reveals that recombination at $V_{OC}$ is essentially SRH recombination (recombination rate follows the minority carrier density profile[37]). Moreover, the back-extracted value of the effective lifetime of the minority charge carriers comes out to be equal to $\tau_{SRH}$ provided in the simulation. Therefore, it can be inferred that other types of recombination mechanisms viz. radiative and Auger recombination, do not play a significant role in defining $V_{OC}$ in these devices. Surprisingly, while the principle of superposition does not follow here (see fig S3 in the supplementary material), we find that $V_{OC}$ values predicted using $n$ and $J_0$ from the dark characteristics near $V = V_{OC}$ in the relation

$$V_{OC} = \frac{nkT}{q}\ln\left(\frac{J_{SC}}{J_0}\right) \quad (4)$$

are in close agreement with the simulated $V_{OC}$.

To further analyze the recombination behavior at $V_{OC}$ in the devices, we plot the percentage contribution of recombination at various regions in fig 2f. It is evident that while the majority of the photogenerated charge carriers were recombined in the active layer in all the cases, significant recombination happened at ETL/LHM interfaces (except for $Sb_2S_3$ solar cell, where both the interfaces and active layers contributed almost equally to the total recombination at $V_{OC}$). Interestingly, we find that despite the similarly deteriorated interface properties, the fraction of charge carrier recombination at the LHM/HTL interface gets reduced drastically with the increase in the percentage of Se in $Sb_2(S_{1-x}Se_x)_3$, and it is mere 0.5% when the percentage of Se is 48%. Nonetheless, the recombination at LHM/ETL interface remains in the range of 20-25%.

*b) Performance optimization routes:*

Now, we move forward to sketch the different routes for performance optimization using the knowledge of the working principle and various loss channels in the devices. Additionally, we discuss the limiting performance of the solar cells based on $Sb_2(S_{1-x}Se_x)_3$ and provide the estimate of the losses through different channels that led to the state of the art performance.

i. Sufficient doping of contact layers:

Equation 2 suggests that the presence of a sufficient electric field in the active layer using heavily doped contact layers would result in net carrier collection, even at a bias equal to $V_{OC}$, and hence $\eta$ (mainly through $V_{OC}$) could be pushed to a higher value. To validate the hypothesis, we perform simulations by varying doping density in the contact layers. The results for $Sb_2S_3$-based solar cells are shown in figure 3. We find that while ETL doping has a little impact on the performance metrics, HTL doping shows a 260 mV change in $V_{OC}$, resulting in a ~3.6% (absolute) shift in $\eta$ for ETL with $10^{18}$ cm$^{-3}$ doping density.

ii. Mobility of the contact layers:

In general, it is believed that contact layers with higher mobility would offer less resistance to the charge carriers[39], and therefore, result in high performance mainly by improving the fill factor ($FF$). To estimate the effect of contact layers mobility on the performance of $Sb_2S_3$ solar cells, we performed simulations using electron mobility in ETL in the range $6.2\times10^{-3}$ to $6.2\times10^3$ cm$^2$/V·s and hole mobility in HTL in the range $2\times10^{-3}$ to 2 cm$^2$/V·s. We observe that contact layer mobility has minimal or no effect on the performance if the contact layers are sufficiently doped. However, in the case of poorly doped contact layers, as expected, the mobility of the contact layers plays a significant role in defining $\eta$. In that context, figure 4 exhibits the simulation results for a doping density of $2\times10^{10}$ cm$^{-3}$ for both HTL and ETL. Approximately 20% (relative) change in FF is evident in the figure, which is translated to ~20% (relative) change in $\eta$.

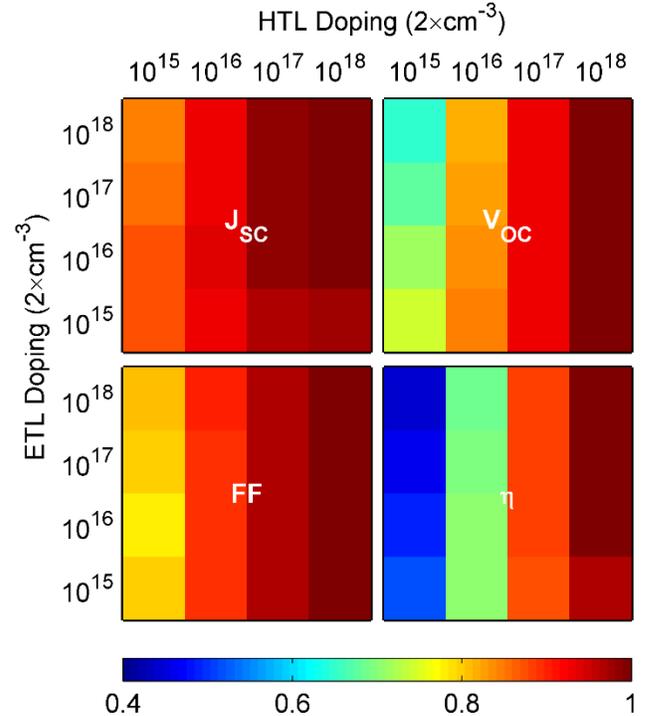

Fig. 3. Effect of contact layers doping on the performance of the $Sb_2S_3$ solar cells. Here, the normalization factor for $\eta$ is $\eta_{norm} = 6.6\%$, $FF$ is $FF_{norm} = 56.5\%$, $J_{SC}$ is $J_{SC_{norm}} = 15.45$ mA/cm$^2$, and $V_{OC}$ is $V_{OC_{norm}} = 758$ mV.

iii. Interface engineering:

If interface engineering[34] is employed to reduce the trap assisted recombination at the interfaces in the device then a commendable increase in $\eta$ could be observed. We have already discussed in figure 2f that interface recombination constitutes more than 20% of the total recombination at $V_{OC}$. It implies that an improvement in the interfacial properties would lead to higher $V_{OC}$ and hence the improved performance. We also notice that the majority of the interfacial loss is observed at ETL/LHM interface rather than HTL/LHM interface (except for $Sb_2S_3$). Therefore, it is necessary to improvise ETL/LHM interface. In figure 5, we show contribution of different losses w.r.t. SQ limiting performance and find that 4-5% (relative) loss in the $\eta$ is due to interface recombination only. Accordingly, if the high-efficiency devices are to be fabricated, it is of the utmost importance to minimize the interface recombination by employing proper interface engineering techniques[34].

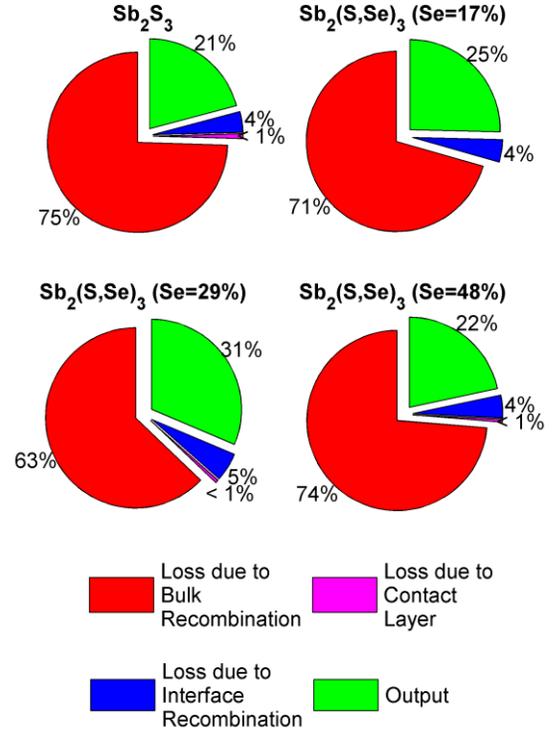

Fig. 5. Contribution of various loss mechanisms in the performance loss in $Sb_2(S,Se)_3$ based solar cells.

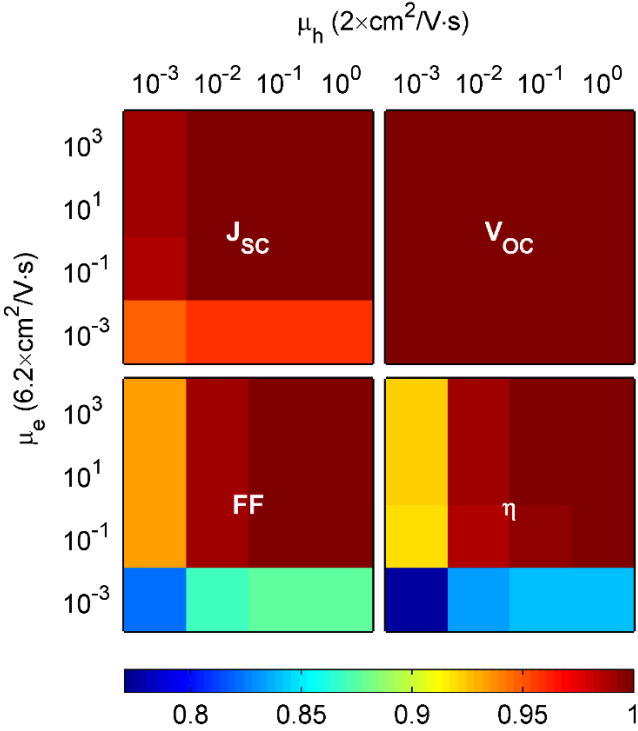

Fig. 4. Effect of charge carriers mobility in the contact layers on the performance of the $Sb_2S_3$ solar cells. Here, the normalization factor for $\eta$ is $\eta_{norm} = 3.5\%$, $FF$ is $FF_{norm} = 47.2\%$, $J_{SC}$ is $J_{SC_{norm}} = 13.4$ mA/cm$^2$, and $V_{OC}$ is $V_{OC_{norm}} = 547$ mV.

iv. Practical performance limit:

While the detailed study of the role of material parameters of $Sb_2(S_{1-x}Se_x)_3$ on the performance characteristics is under investigation, we are eager to discuss the limiting performance and various losses in these devices in the paper for the completeness as well as to provide design directions to the researchers in the field. Detailed balance of the photogenerated charge carriers suggest that the limiting performance corresponding to Se percentage in $Sb_2(S_{1-x}Se_x)_3$ being 0%, 17%, 29%, and 48% are 28.8%, 31.1%, 32.2%, and 32.7%, respectively[40]. However, state of the art devices have only 20-30% efficiency of that. We find that the major part of this huge loss is due to poor charge carrier lifetime in the bulk of the material as indicated by 60–75% (relative) loss in the performance w.r.t. SQ limit due to recombination in the bulk of the LHM (see fig 5). Although we find that contact layers are almost optimum and do not contribute to any significant loss, approximately 5% (relative) increase in $\eta$ could be achieved by proper interface engineering. Hence, the practical efficiency limit for the devices under investigation using the active layers with the state of the art material characteristics are 25-35% of the corresponding SQ limit (7-11% absolute value). It should be noticed here that 11% $\eta$ is in fact the practical efficiency limit (as of now) for the devices under investigation. However, the fabrication of the active layers with the reduced traps would indeed reduce the red area in the pie charts and result in further improvement in that limit and the realizable performance.

## IV CONCLUSIONS

To summarize, here, we have identified parameter space for $Sb_2(S_{1-x}Se_x)_3$ solar cells and calibrated the same using experimental current-voltage characteristics reported in the literature. The modeling strategy is then extended to identify different schemes that could potentially improve the state of the art performance of the device. We find that while ETL doping has negligible to zero effect on the performance, insufficient doping of HTL could lead to a 60% (relative) loss in the performance. Furthermore, the mobility of the charge carriers in the contact layers play a little role if the layers are moderate to heavily doped; however, deficient doping indeed leads to mobility dependent performance, and a spread of 20% (relative) in the efficiency has been observed for hole mobility in HTL in the range of $2\times10^{-3}$ to 2 cm$^2$/V·s if the doping density in the contact layers is of the order of $10^{10}$ cm$^{-3}$. In addition, we elucidated that interface engineering is of utmost importance if the high-efficiency devices have to be fabricated. We also show that without any further improvisation in the characteristics of the active layer, ~11% efficient solar cell using Sb$_2$(S$_{0.71}$Se$_{0.29}$)$_3$ can be fabricated by employing the schemes discussed in the paper, although the SQ limit for the same is ~32.2%. Any further improvement is only possible by the improvisation of material parameters of the active layer. It is worth mentioning here that a significant loss in terms of $J_{SC}$ has been observed in these devices, which could be mitigated by thickness optimization, and the same is under investigation. Conclusively, with the state of the art material properties, antimony chalcogenide have shallow potential (due to poor performance) to have a significant market share in the solar cell technology; however, if the due emphasis is given towards material engineering of this class of material to prepare solar-grade LHM, then it could provide efficient and stable solar cells.